\documentclass{aa}  

\usepackage{graphicx}
\usepackage{txfonts}
\usepackage{color} % for text colors
\usepackage{xspace}

\begin{document}

   \title{Outflows in the presence of cosmic rays and waves with cooling}

   %\subtitle{I. Overviewing the Cooling-mechanism}

   \author{C. M. Ko
          \inst{1,2}
         \and
         %\fnmsep
          B. Ramzan\inst{1}
          \and
          D. O. Chernyshov
          \inst{3,4}
          }

   \institute{Graduate Institute of Astronomy, National Central University, Zhongli Dist., Taoyuan City 32001, Taiwan, (R.O.C.)\\
              %\email {cmko@astro.ncu.edu.tw}
              %\email{bilal@gm.astro.ncu.edu.tw}
              \and
             Department of physics, National Central University, Zhongli Dist., Taoyuan City 32001, Taiwan, (R.O.C.)\\
              \and
               I. E. Tamm Theoretical Physics Division of P. N. Lebedev Institute of Physics, Leninskii pr. 53, 119991 Moscow, Russia\\
             \and
             Moscow Institute of Physics and Technology (State University), 9, Institutsky lane, Dolgoprudny, 141707, Russia\\
            %\email{chernyshov@dgap.mipt.ru}
             \email {cmko@astro.ncu.edu.tw},
             \email{bilal@astro.ncu.edu.tw}
            %\thanks{
             %\email cmko@astro.ncu.edu.tw}
             %\thanks{
                     %d1049601@gm.astro.ncu.edu.tw}
             }

  % \date{Received February 00, 2021; accepted March 00, 2021}

% \abstract{}{}{}{}{} 
% 5 {} token are mandatory
 
  \abstract
   {Plasma outflow from a gravitational potential well with cosmic rays and self-excited Alfv\'en  waves with cooling and wave damping is studied in the hydrodynamics regime.}
   {We study outflows in the presence of cosmic ray and Alfv\'en waves including the effect of cooling and wave damping. We seek physically allowable steady-state subsonic-supersonic transonic solutions.
   }
   {We adopted a multi-fluid hydrodynamical model for the cosmic ray plasma system. Thermal plasma, cosmic rays, and self-excited Alfv\'en  waves are treated as fluids. Interactions such as cosmic-ray streaming instability, cooling, and wave damping were fully taken into account. We considered one-dimensional geometry and explored steady-state solutions. The model is reduced to a set of ordinary differential equations, which we solved for subsonic-supersonic transonic solutions with given boundary conditions at the base of the gravitational potential well.}
   {We find that physically allowable subsonic-supersonic transonic solutions exist for a wide range of parameters. We studied the three-fluid system (considering only forward-propagating Alfv\'en waves) in detail. We examined the cases with and without cosmic ray diffusion separately. Comparisons of solutions with and without cooling and with and without wave damping  for the same set of boundary conditions (on density, pressures of thermal gas, cosmic rays and waves) are presented. We also present the interesting case of a four-fluid system (both forward- and backward-propagating Alfv\'en waves are included), highlighting the intriguing relation between different components.}
   {}


 \maketitle
 
\section{Introduction}

Cosmic ray energy density and the energy density of other components of the interstellar medium
(e.g., different phases of gas, magnetic field) are on the same order of magnitude
\citep[e.g.,][]{Ferrire_2001,Cox_2005}.
The equipartition of energy between the different components indicates there are significant interactions among these components of the interstellar medium.
Cosmic rays can play a dynamical role in the structure and evolution of the interstellar as well as intergalactic medium.
For instances, from the hydrodynamic perspective, cosmic rays can affect instabilities, such as, Parker instability, Jeans instability, magnetorotional instability
\citep[e.g.,][]{Parker_1969,Parker_1966,Hanasz_2000,Hanasz_2003,Ryu_2003,Kuwabara_2004,Kuwabara_2006,Ko_2009,Lo_2011,Kuwabara_2015,Heintz_2018,Heintz_2020,Kuwabara_2020},
and they can modify structures and outflows 
\citep[e.g.,][]{Koetal_1991,Yang_2012,Girichidis_2016,Dorfi_2012,Recchia_2016,Ruszkowski_2017,Mao_2018,Farber_2018,Holguin_2019,Dorfi_2019,Yu_2020,Recchia_2020,Ramzan_2020}.

The cosmic-ray driven wind mechanism was first discussed by \citet{Ipavich_1975}.
Subsequently, \citet{Breitschwerdt_1991,Breitschwerdt_1993},
and
\citet{Zirakashvili_1996}, laid down the proper formulation of cosmic ray propagation with diffusion and wave damping for cosmic-ray driven winds.
Later studies 
\citep[e.g.,][]{Everett_2008,Recchia_2016,Wiener_2017,Ruszkowski_2017,Farber_2018} showed that the continuous gas outflows from galaxies can affect their evolution directly. For more details, we refer to the works of \citet{Zhang_2018} and \citet{Recchia_2020}.

Steady-state thermal wind transonic solutions with cooling for a wide range of mass-loading factors, energy-loading factors, galaxy mass, and galaxy radius was studied by \citet{Bustard_2016}. These authors showed that radiative losses or cooling can enormously affect the wind solutions. The cooling of the gas is important for finding smooth transition from the subsonic to supersonic branch (transonic solutions). 
\citet{Bustard_2016} presented the cases of thermal driven wind with cooling and provided feasible transonic solutions. 
In this article, we investigate the effect of cooling on outflows under the dynamical effect of cosmic rays and waves.
We explore different possibilities of subsonic-supersonic transonic solutions when cosmic ray diffusion and wave damping mechanism are considered.
We note that the terms ``subsonic'' and ``supersonic'' in this article do not refer to the propagation speed of a disturbance in the medium, but rather the speed that is determined by the critical point condition of the system. We call this speed the ``effective sound speed.''

The paper is organized as follows.
In Section~\ref{sec:model}, we present the four-fluid model of the cosmic-ray-plasma system in flux-tube formation.
The four fluids are thermal plasma, cosmic ray, forward- and backward-propagating Alfv\'en waves.
A simpler system, namely, a system with only a forward-propagating wave (i.e., a three-fluid system or a one-wave system) is studied in Section~\ref{sec:withandwithout_diffusion} and outflows with and without cosmic ray diffusion are discussed in detail. 
Section~\ref{sec:4fluid-wdiff} describes an example of the four-fluid system.
In Section~\ref{sec:Summary}, we present our remarks and a summary.
%%%%%%%%%%%%%%%%%%%%%%%%%%%%%%%%%%%%%%%%%%%%%%%%%%%%%%%%%%%%%%
\section{Hydrodynamical model for a cosmic ray-plasma system}
\label{sec:model}
Cosmic rays interact with magnetised plasma via the embedded magnetic field, in particular, magnetic irregularities or waves. As a result, cosmic rays advect with and get diffused through the plasma. As cosmic rays stream through the plasma, they excite hydromagnetic waves via streaming instabilities. When the energy density of cosmic rays (and waves) are on the same order as that of the thermal plasma, their feedback onto the thermal plasma cannot be neglected. Energy exchange is facilitated by various pressure gradients and wave damping. To study the dynamics of the cosmic ray plasma system, we adopted the four-fluid hydrodynamical model (or two-wave system) put forward by \citep{Ko_1992} (see also Zweibel 2017). In this model, the thermal plasma, cosmic rays, and two Alfv\'en waves (one propagating forward along the magnetic field and one propagating backward) are treated as fluids. In the flux-tube formulation (i.e., considering the dynamics along the magnetic field only) and for the steady state, the governing equations are \citep{Ramzan_2020}:
\begin{equation}\label{eq:bfield}
  B \Delta = \psi_B\,,
\end{equation}
\begin{equation}\label{eq:mass}
  \rho U \Delta = \psi_m\,,
\end{equation}
\begin{equation}\label{eq:momentum}
  \rho U\frac{d U}{d\xi} = -\frac{d}{d\xi}\left(P_g+P_c+P^+_w+P^-_w\right)+\rho g_e\,,
\end{equation}
\begin{equation}\label{eq:gas}
  \frac{1}{\Delta}\frac{d F_g \Delta}{d\xi} = U\frac{d P_g}{d\xi} +L^+_w+L^-_w-\Gamma \,,
\end{equation}
\begin{equation}\label{eq:cr}
  \frac{1}{\Delta}\frac{d F_c \Delta}{d\xi} = \left[U+(e_+-e_-)V_A\right]\frac{d P_c}{d\xi}+\frac{P_c}{\tau}\,,
\end{equation}
\begin{equation}\label{eq:waves}
  \frac{1}{\Delta}\frac{d F^\pm_w \Delta}{d\xi} = U\frac{d P^\pm_w}{d\xi}\mp e_\pm V_A\frac{d P_c}{d\xi} - \frac{P_c}{2\tau} - L^\pm_w\,,
\end{equation}
where the energy fluxes are
\begin{equation}\label{eq:gflux}
  F_g = \left(E_g+P_g\right)U = \frac{\gamma_g P_g}{\left(\gamma_g-1\right)}\,U\,,
\end{equation}
\begin{align} \label{eq:crflux}
  F_c & = \left(E_c+P_c\right)\left[U+(e_+-e_-)V_A\right]- D_c \nonumber \\
      & = \frac{\gamma_c P_c}{\left(\gamma_c-1\right)}\left[U+(e_+-e_-)V_A\right]-\frac{\kappa}{\left(\gamma_c-1\right)}\frac{d P_c}{d\xi} \,,
\end{align}\label{eq:wflux}
\begin{equation}
  F^\pm_w = E^\pm_w\left(U\pm V_A\right)+P^\pm_w U = P^\pm_w\left(3U\pm 2V_A\right)\,.
\end{equation}
Equation~(\ref{eq:bfield}) and Equation~(\ref{eq:mass}) come from the divergent free magnetic field and mass continuity equation,
$\psi_B$ and $\psi_m$ are called the magnetic flux and the mass flow rate, respectively;
$\Delta(\xi)$ is the cross-sectional area of the magnetic flux tube and
$\xi$ is the coordinate along the magnetic field.
In Equation~(\ref{eq:momentum}),  
$\rho$ and $U$ are the density and flow velocity of the plasma,
$P_g$, $P_c$, and $P^\pm_w$ are the pressures of the thermal plasma, cosmic rays, and waves ($\pm$ denote forward- and backward-propagating Alfv\'en waves).
Also, $g_e$ is the external gravitational field.
Equation~(\ref{eq:gas}) describes the influence of gas by cooling, $\Gamma$, and by heating via wave damping, $L^\pm_w$.
The terms $\mp e_\pm V_A dP_c/d\xi$ in Equations~(\ref{eq:cr}) and ~(\ref{eq:waves}) are wave excitation by cosmic ray streaming instability.
In Equation~(\ref{eq:crflux}), 
$\kappa$ is the diffusion coefficient of cosmic rays and $D_c=\kappa d E_c/d\xi$ is the cosmic ray diffusive flux,
$P_c/\tau$ represents stochastic acceleration.
Also, $V_A= B /\sqrt{\mu_0\rho}$ is the Alfv\'en speed. We adopted the following working model for $e_\pm$, $\kappa$ and $1/\tau$
\citep[e.g.,][]{Skilling,Ko_1992}:
\begin{align}\label{eq:epmkappatau}
  e_\pm & = \frac{\nu_\pm}{(\nu_++\nu_-)}\,, \nonumber \\
  \kappa & = \frac{c^2}{3(\nu_++\nu_-)}\,, \\
  \frac{1}{\tau} & = \frac{16 \nu_+\nu_- V_A^2}{(\nu_++\nu_-)c^2}\,, \nonumber
\end{align}
where $\nu_\pm$ are the collision frequencies of cosmic rays by forward- and backward-propagating waves.
We take $\nu_\pm = c^2 P^\pm_w/ \eta$ in the calculations in Sections 3 and 4. The smaller the $\eta$, the stronger the coupling strength.

%%%%%%%%%%%%%%%%%%%%%%%%%%%%%%%%%%%%%%%%%%%%%%%%%%%%%%%%%%%%%%
\section{Wind equation in constant flux-tube geometry}
\label{sec:two_wave_diffusion}
In the following sections, we focus on outflows (especially transonic outflows) under the influence of gas cooling, wave damping, and cosmic ray diffusion as well. To avoid complications arising from a divergent flux-tube, we restrict our discussions to constant flux-tube geometry only, that is, we take $\Delta$ to be a constant. To study outflows, it is useful to express the momentum equation (Equation~(\ref{eq:momentum})) in the form of a wind equation (cf. the classic stellar wind in \citet{Parker_1958}),

\begin{align}\label{eq:two_wave_wind}
  & \left(1- M^{-2}_{\rm eff}\right)U\frac{dU}{d\xi} = \nonumber \\
  &\quad g_e-\frac{1}{\rho}\frac{d P_c}{d\xi}
  \left[\frac{e_+(1+\frac{1}{2}M_A^{-1})}{(1+M_A^{-1})}+\frac{e_-(1-\frac{1}{2}M_A^{-1})}{(1-M_A^{-1})}\right] \nonumber \\
  &\quad +\frac{1}{\rho U}\left[\frac{P_c}{2(1-M_A^{-2})\tau}-(\gamma_g-1)\left(L^+_w+L^-_w-\Gamma \right) \right. \nonumber \\
  &\quad\quad\quad \left. +\frac{L^+_w}{2(1+M_A^{-1})}+\frac{L^-_w}{2(1-M_A^{-1})}\right]\,.
\end{align}
where $M_A= \sqrt{\mu_0} \psi_m/(\psi_B\sqrt{\rho})=1/({\tilde\psi\sqrt{\rho}})$ is the Alfv\'en Mach number and $M_{\rm eff}=U/a_{\rm eff}$ is the ``effective Mach number.'' $d P_c/d\xi$ appears explicitly only if cosmic ray diffusion coefficient is non-zero. The case without cosmic ray diffusion is discussed in Section~\ref{sec:without_diffusion}.
Here the ``effective sound speed'' is:
\begin{equation}\label{eq:sound_speed}
  a^2_{\rm eff} = a^2_g+a^{+\,2}_w\frac{\left(1+\frac{1}{3}M_A^{-1}\right)}{\left(1+M_A^{-1}\right)}
  +a^{-\,2}_w\frac{\left(1-\frac{1}{3}M_A^{-1}\right)}{\left(1-M_A^{-1}\right)}\,,
\end{equation}
where
\begin{equation}\label{eq:sound_speeds}
  a^2_g = \frac{\gamma_g P_g}{\rho}\,,
  \quad a^{+\,2}_w = \frac{3 P^+_w}{2\rho}\,,
  \quad a^{-\,2}_w = \frac{3 P^-_w}{2\rho}\,.
\end{equation}
We note that the ``effective sound speed'' is not necessarily the speed of the propagation of a disturbance in the medium. It is determined by the ``critical point'' of the system, as seen in the term on the left-hand side of Equation~(\ref{eq:two_wave_wind}); more details are given in Appendix~\ref{sec:critical_behaviour}. 

Equations for the gas, cosmic rays, and waves are:
\begin{equation}\label{eq:two_wave_gas}
  \frac{d P_g}{d\xi} = -\frac{\gamma_g P_g}{U}\frac{dU}{d\xi}+\frac{(\gamma_g-1)}{U}\left(L^+_w+L^-_w-\Gamma \right)\,,
\end{equation}

\begin{align}\label{eq:two_wave_wave}
  & \frac{d P^\pm_w}{d\xi} = -\frac{3 P^\pm_w}{2U}\frac{(1\pm\frac{1}{3}M_A^{-1})}{(1\pm M_A^{-1})}\frac{dU}{d\xi}
  \mp\frac{e_\pm M_A^{-1}}{2(1\pm M_A^{-1})}\frac{d P_c}{d\xi} \nonumber \\
  &\quad -\frac{P_c}{4U(1\pm M_A^{-1})\tau}-\frac{L^\pm_w}{2U(1\pm M_A^{-1})}\,,
\end{align}
\begin{align}\label{eq:two_wave_cr}
  & \frac{d P_c}{d\xi} = -\frac{\gamma_c P_c}{U}\frac{[1+\frac{1}{2}(e_+-e_-)M_A^{-1}]}{[1+(e_+-e_-)M_A^{-1}]}\frac{dU}{d\xi} \nonumber \\
& \quad\quad\quad -\frac{\gamma_c P_c M_A^{-1}}{[1+(e_+-e_-)M_A^{-1}]} *\frac{d}{d\xi}(e_+-e_-)  \nonumber \\
&\quad\quad\quad +\frac{(\gamma_c-1) P_c}{U[1+(e_+-e_-)M_A^{-1}]\tau} \nonumber \\
&\quad\quad\quad +\frac{1}{U[1+(e_+-e_-)M_A^{-1}]}\frac{d}{d\xi}\left(\kappa\frac{dP_c}{d\xi}\right)\,.
\end{align}
In the following discussion, we take the galactic wind as our fiducial example of outflows. As a working model, we adopted non-linear Landau damping as the wave damping mechanism, $L_w^\pm \propto P_w^{\pm 2}\sqrt{P_g/\rho B^2}$, and bremsstrahlung radiation for cooling mechanism,  $\Gamma \propto \rho^2\sqrt{P_g/\rho}$. The corresponding proportionality constants are denoted by $\delta^\pm$ and $\delta_c$ in the following calculations. The gravitational field is $g_e=-GM/(\xi+a)^2$ (cf. a Kuzmin-like field near the center of the disk), where $a$ is the characteristic scale of the field.
%%%%%%%%%%%%%%%%%%%%%%%%%%%%%%%%%%%%%%%%%%%%%%%%%%%%%%%%%%%%%%
\section{Transonic outflows in a three-fluid system}
\label{sec:withandwithout_diffusion}
In this section, we consider a simpler system, namely, a three-fluid or one-wave system. In this system, only the forward-propagating wave is considered,
namely, $e_+=1$, $e_-=0$, $1/\tau=0$.
We seek physically allowable solutions from a given set of boundary conditions at the base of the potential well. Physically allowable solution means the outflow quantities are single-value with respect to $\xi$, and the velocity and pressures are non-negative and finite. In particular, we are interested in subsonic-supersonic transonic solutions.
Here, we discuss two different cases: (1) systems without cosmic ray diffusion and (2) systems with cosmic-ray diffusion. For subsonic or supersonic solutions, these two cases are not expected be too different \citep{Ramzan_2020}. However, for transonic solutions, they can be very different because the effective sound speeds for the two cases are very different (cf. Equation~(\ref{eq:sound_speed}) and ~(\ref{eq:sound_speed_1})).
%%%%%%%%%%%%%%%%%%%%%%%%%%%%%%%%%%%%%%%%%%%%%%%%%%%%%%%%%%%%%%
\subsection{Three-fluid outflow without cosmic ray diffusion}
\label{sec:without_diffusion}
In this subsection, we study three-fluid outflows by neglecting cosmic ray diffusion (i.e., cosmic ray is strongly coupled to thermal gas). Setting $e_+=1$, $e_-=0$, $1/\tau=0$ and $\kappa=0$ in Equation~(\ref{eq:two_wave_cr}), the cosmic-ray pressure can be expressed analytically as Equation~(\ref{eq:pc_no_diff}) (or see Ramzan et al. 2020) and the three-fluid wind equation becomes (cf. Equation~(\ref{eq:two_wave_wind})):
\begin{align}\label{eq:one_wave_wind}
   \left(1- M^{-2}_{\rm eff}\right)U\frac{dU}{d\xi} = g_e +(\gamma_g-1) \frac{\Gamma}{\rho U} \\ \nonumber
  +\left[\frac{1}{2(1+M_A^{-1})}-(\gamma_g-1)\right] \frac{L_w^+}{\rho U}\,,
\end{align}
where the ``effective sound speed'' is modified to (cf. Equation~(\ref{eq:sound_speed})):
\begin{equation}\label{eq:sound_speed_1}
  a^2_{\rm eff} = a^2_g+a^{+\,2}_w\frac{\left(1+\frac{1}{3}M_A^{-1}\right)}{\left(1+M_A^{-1}\right)}
  +a^{\,2}_c\frac{\left(1+\frac{1}{2}M_A^{-1}\right)^2}{\left(1+M_A^{-1}\right)^2}\,,
\end{equation}
and $a^{2}_c = \gamma_c P_c/\rho$.
We explore outflow against gravity, thus, $g_e$ is negative (and approaches zero at large distances away from the gravitational potential well). For convenience, we call the right hand side of the wind equation (for this subsection it is Equation~(\ref{eq:one_wave_wind})) the driving term. If there is no cooling and no wave damping, then the driving term is always negative. Thus, in the subsonic regime ($M_{\rm eff}<1$), $dU/d\xi>0$ and the outflow speed is monotonically increasing with $\xi$. We may call this accelerating outflow. 
There are two possibilities: (1) the accelerating outflow is always subsonic and approaches an asymptotic value at large distances or (2) the flow speed of the solution approaches the effective sound speed at finite distance, then $dU/d\xi\rightarrow\infty$ and the solution becomes unphysical. By the same token, in the supersonic regime ($M_{\rm eff}>1$), $dU/d\xi<0$, and the outflow is a decelerating outflow. Here, there are also two possibilities: (1) a decelerating supersonic outflow that approaches a finite velocity at large distances or (2) the solution approaches the effective sound speed and becomes unphysical. 

If cooling is included (but still ignoring wave damping), then the driving term can be positive or negative. Thus, a new possibility arises. It is then possible to have a solution that as its flow speed approaches the effective sound speed and driving term also approaches zero simultaneously. $dU/d\xi$ can be finite at this point (the sonic point or critical point), and the solution can pass through the critical point. A transonic solution is born. A more detailed analysis of this possibility can be found in Appendix~\ref{sec:3f_without_diffusion}. 

When wave damping is included, its contribution to the driving term can be positive or negative. For sub-Alfv\'enic flow the contribution is always negative if $\gamma_g>5/4$, while for super-Alfv\'enic flow the contribution is always negative if $\gamma_g>3/2$ and always positive if $\gamma_g<5/4$.
In this work we only consider cases with $\gamma_g=5/3$. 
We seek physically allowable solutions to the wind equation and pressure equations (Equations (14)-(17)) subjected to a set of boundary conditions of $(U,P_g,P_c,P_w^+,\rho,B)$ at the base of the gravitational potential well. 
However, not every set of boundary conditions gives us a physically allowable solution. Physically allowable solutions are divided into subsonic solutions, supersonic solutions, and transonic solutions. For the purposes of this study, we are interested in subsonic-supersonic transonic solutions only. At the outset we chose a set $(P_{gb}, P_{cb}, P_{wb}^+, \rho_{b}, B_{b})$ at the base of the potential well and then adjusted $U_{b}$ to obtain the transonic solution.

\begin{figure*}[htb]
\centering
\vskip 0.5cm
\includegraphics[width=0.8\textwidth]{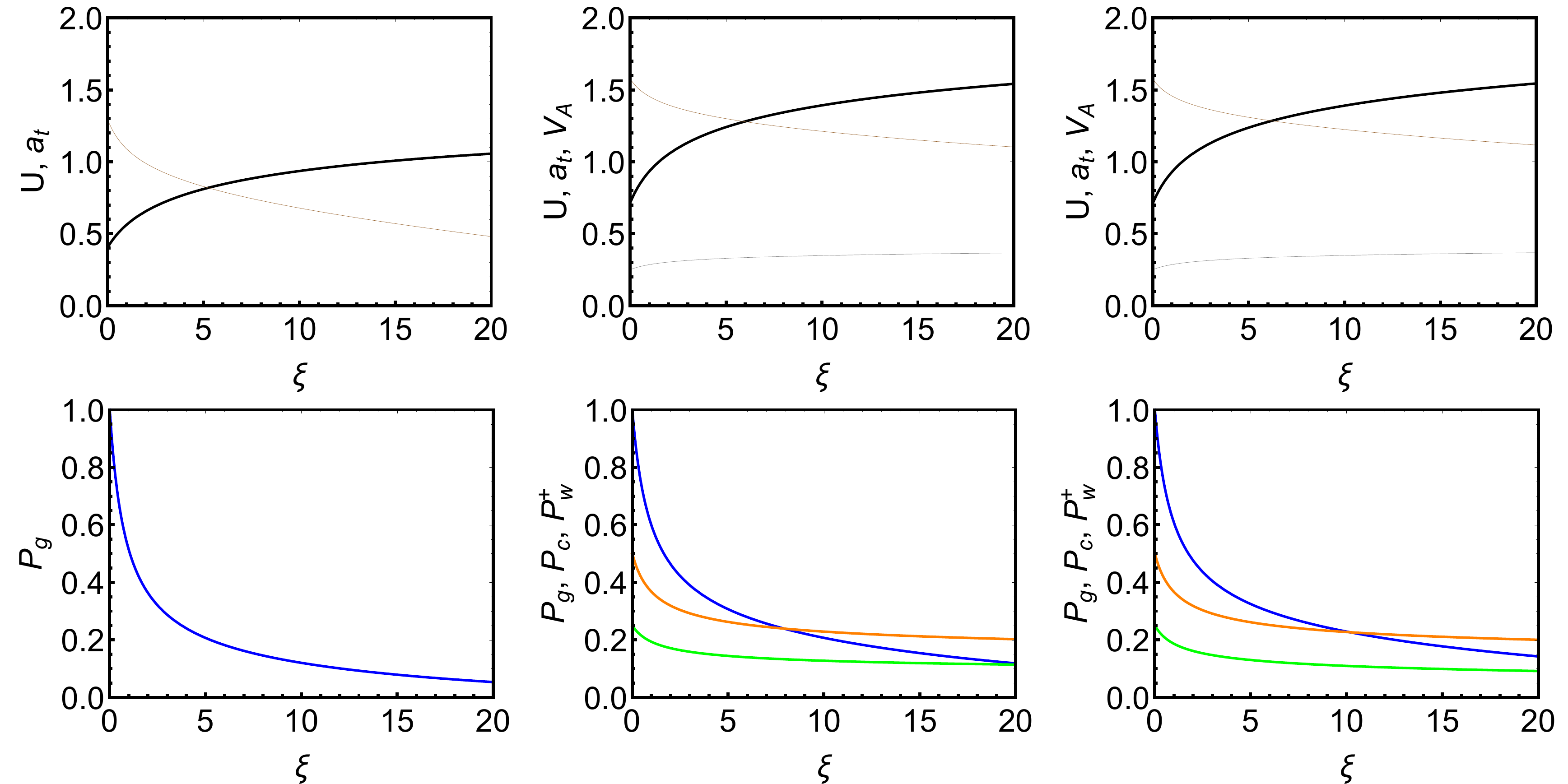}
\vskip 0.5cm
\caption{
Transonic solutions of a three-fluid system without cosmic ray diffusion. The transonic solution is subsonic-supersonic transition. In this figure, $\gamma_g=5/3$, $\gamma_c=4/3$, $GM=1$, $a=1$.
{\it Left column}: Pure thermal outflow with cooling ($\delta_c=0.1$).
{\it Middle column}: Three-fluid outflow with cooling ($\delta_c=0.1$) but without wave damping ($\delta^+=0$).
{\it Right column}: Three-fluid outflow with cooling ($\delta_c=0.1$) and wave damping ($\delta^+=0.1$). 
{\it Upper row}: Black line and thin brown lines show the flow speed and the effective sound speed of the system. For the three-fluid cases, Alfv\'en speed is also shown (thin gray lines).
{\it Lower row}: Blue, orange, and green lines show the pressures of the thermal gas, cosmic ray, and forward-propagating wave, respectively. The boundary conditions $(P_{gb},\rho_b,B_b)$ are the same for all three cases, and $(P_{cb},P_{wb}^+)$ are the same for the two three-fluid cases. In this example, $P_{gb}=1$, $P_{cb}=0.5$, $P_{wb}^+=0.25$, $\rho_b=1$, $B_b=\psi_B^\prime=0.25$. The boundary condition $U_b$ is adjusted to obtain the transonic solutions.
}
\label{fig:3-fluid-nodiff}
\end{figure*}

In Figure~\ref{fig:3-fluid-nodiff}, we show a typical result of three-fluid outflows with cooling. We present different cases side by side. The middle column is the case without wave damping and the right column is the case with wave damping. The boundary conditions at the base $(P_{gb},P_{cb},P_{wb}^+,\rho_b,B_b)$ for both cases are the same, but $U_b$ are different in order to obtain a transonic solution for the two cases. As shown in the figure, the velocity and pressure profiles of the two cases are quite similar. We also included a pure thermal outflow for comparison (the left column). The boundary condition for the pure thermal outflow has the same $(P_{gb}, \rho_b,B_b)$ as the three-fluid cases, while $U_b$ is adjusted to get the transonic solution. The velocity of the pure thermal outflow is significantly smaller than the three-fluid cases (black solid line in the upper row of Figure~\ref{fig:3-fluid-nodiff} is flow speed). This is expected as the corresponding effective sound speeds are different (thin brown line in the upper row of Figure~\ref{fig:3-fluid-nodiff} is effective sound speed). The effective sound speed of the three-fluid cases has extra contributions from waves and cosmic rays (see  Equation~(\ref{eq:sound_speed_1})). Nevertheless, the thermal gas pressure profiles of the three cases are shown to be rather similar (see the lower row of Figure~\ref{fig:3-fluid-nodiff}).
 %%%%%%%%%%%%%%%%%%%%%%%%%%%%%%%%%%%%%%%%%%%%%%%%%%%%%%%%%%%%%
%%%%%%%%%%%%%%%%%%%%%%%%%%%%%%%%%%%%%%%%%%%%%%%%%%%%%%%%%%%%%%
\subsection{Three-fluid outflow with cosmic ray diffusion}
\label{sec:3fluid-wdiff}
An important feature of cosmic ray transport is the diffusion of cosmic rays through the thermal plasma. The diffusion coefficient indicates the coupling of cosmic rays with the plasma; the larger the diffusion coefficient, the weaker the coupling. We modeled the diffusion coefficient as inversely proportional to the wave pressure (thus, as the wave pressure tends to zero, cosmic rays tend to be decoupled from the plasma). 
The wind equation is Equation~(\ref{eq:two_wave_wind}) with only the forward-propagating wave, and $e_+=1$, $e_-=0$, $1/\tau=0$. We note that cosmic ray pressure gradient appears explicitly in the driving term of the wind equation (the right hand side of Equation~(\ref{eq:two_wave_wind})).
In comparison with the cases without cosmic ray diffusion (Section~\ref{sec:without_diffusion}), the presence of cosmic ray diffusion shifts the narrative of the existence of transonic solution in two ways: the cosmic ray contribution to the effective sound speed disappears (see Equation~(\ref{eq:sound_speed})), and the role of cosmic rays in the driving term. In fact, transonic solution can exist in three-fluid system with diffusion even without cooling and wave damping (see Appendix~\ref{sec:3f_with_diffusion_withoutcooling}). However, we have opted to focus on the role of cooling on transonic solution in the present study.

\begin{figure*}[htb]
\centering
\vskip 0.5cm
\includegraphics[width=0.8\textwidth]{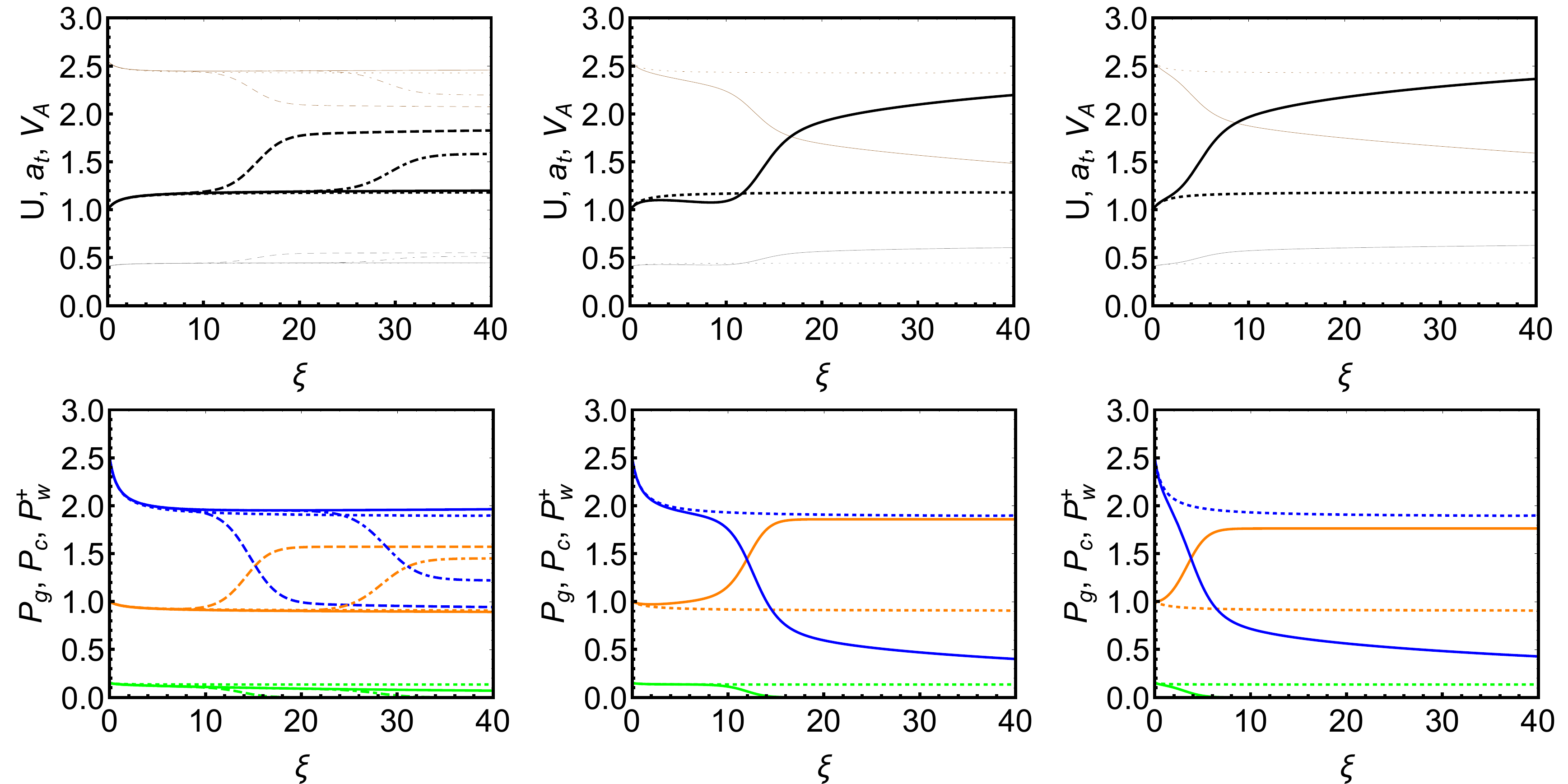}
\vskip 0.5cm
\caption{
Comparison of transonic and subsonic solutions of a three-fluid system with cosmic ray diffusion. In this figure, $\gamma_g=5/3$, $\gamma_c=4/3$, $GM=1$, $a=1$, $\eta=1$.
Cases with wave damping and/or cooling are compared to the cases without both wave damping and cooling.
All cases have the same boundary conditions $(U_b,P_{gb},P_{cb},P_{wb}^+,\rho_b,B_b)$ at the base except $D_{cb}$.
The diffusive flux (or cosmic ray pressure gradient) at the base is adjusted to get physically allowable solutions (subsonic solutions for the cases without cooling and transonic solution for the cases with cooling).
In this example, $U_b=1$, $P_{gb}=2.5$, $P_{cb}=1$, $P_{wb}^+=0.15$, $\rho_b=2/3$, $B_b=\psi_B^\prime=1/3$.
Dotted line displays the case in which both wave damping and cooling are “OFF” ($\delta_c=0$, $\delta^+=0$).
{\it Left column}: Dashed, dot-dashed, and solid lines are the cases in which only wave damping is “ON” ($\delta_c=0$, $\delta^+=0.1$). 
The three solutions correspond to slightly different $D_{cb}$ (see text).
{\it Middle column}: Solid line is the case with only cooling is “ON” ($\delta_c=0.1$, $\delta^+=0$).
{\it Right column}: Solid line is the case in which both  cooling and wave damping are “ON” ($\delta_c=0.1$, $\delta^+=0.1$).
{\it Upper row}: Black, thin brown, and grey lines show the flow speed, the effective sound speed and the Alfv\'en speed, respectively.
{\it Lower row}: Blue, orange, and green lines show the pressures of the thermal gas, cosmic ray, and forward-propagating wave, respectively.
}
\label{fig:3fluid_diff}
\end{figure*}

We explore solutions for the wind equation (Equation~(\ref{eq:two_wave_wind})), together with the corresponding pressure equations, namely, Equations (14) to (16) that are subject to boundary conditions at the base of the gravitational potential well. In addition to $(U_b,P_{gb},P_{cb},P_{wb}^+,\rho_b,B_b)$, we need one more boundary condition for the gradient of cosmic-ray pressure (or cosmic-ray diffusive flux $D_{cb}$) at
the base.
In Figure~\ref{fig:3fluid_diff}, we compare the case with both cooling and
wave damping are ``OFF'' to cases where wave damping is ``ON’’ or cooling is ``ON,’’ or both are ``ON.'' In this example, wave damping is ``OFF'' and ``ON,'' referring to $\delta^+=0$ and 0.1 and cooling is ``OFF'' and ``ON,’’ referring to $\delta_c=0$ and 0.1. All cases have the same boundary conditions $(U_b,P_{gb},P_{cb},P_{wb}^+,\rho_b,B_b)$ except $D_{cb}$. For those for which cooling is ``ON,’’ we adjust $D_{cb}$ to obtain the transonic solution (subsonic-supersonic transition), while for those for which cooling is ``OFF,’’ we again adjust $D_{cb}$ to obtain the physically allowable solution (i.e., subsonic solution that extends to large distances).

In Figure~\ref{fig:3fluid_diff}, the upper row shows the outflow speed (black), effective sound speed (thin brown) and Alfv\'en speed (thin gray). The lower row shows the thermal pressure (blue), cosmic ray pressure (orange) and wave pressure (green).
The case where both wave damping and cooling are ``OFF’’ is compared to the case that only wave damping is ``ON’’ in the left column, as well as to the case that only cooling is ``ON’’ in the middle column, and to the case that
both wave damping and cooling are ``ON’’ in the right column.

In the left column, all the cases are without cooling and they all belong to the subsonic outflow. The dotted line is the case without wave damping and cooling. The dashed, dot-dashed, and solid lines belong to wave damping case and all with the same $\delta^+ = 0.1$, but with slightly different $D_{cb}$. Here, $D_{cb}$ decreases from dashed to dot-dashed to solid and the solid line has the minimum $D_{cb}$ beyond which there is no physically allowable solution in this example. The solid line is barely different from the dotted line. The dashed and dot-dashed lines have a rapid increase in $U$ and $P_c$ (and a rapid decrease in $P_g$) somewhere along the flow where $P_w^+$ becomes diminish and cosmic rays decouple from the thermal gas. These are typical quasi-thermal outflows discussed in \citet{Ramzan_2020}. Appendix~\ref{sec:Behaviorofpressures} gives a summary of different flow profiles in three-fluid system (with forward-propagating wave).

In the middle and the right columns, cooling is included and the solutions are subsonic-supersonic transonic flows (solid lines). The profiles carry the flavor of quasi-thermal outflows. It appears that the transition (sonic point) takes place at a further distance for the case without wave damping (the middle column). However, the profiles at large distances for both cooling cases are quite similar in this example.

\begin{figure}[htb]
\centering
\vskip 0.8cm
\includegraphics[width=0.5\textwidth]{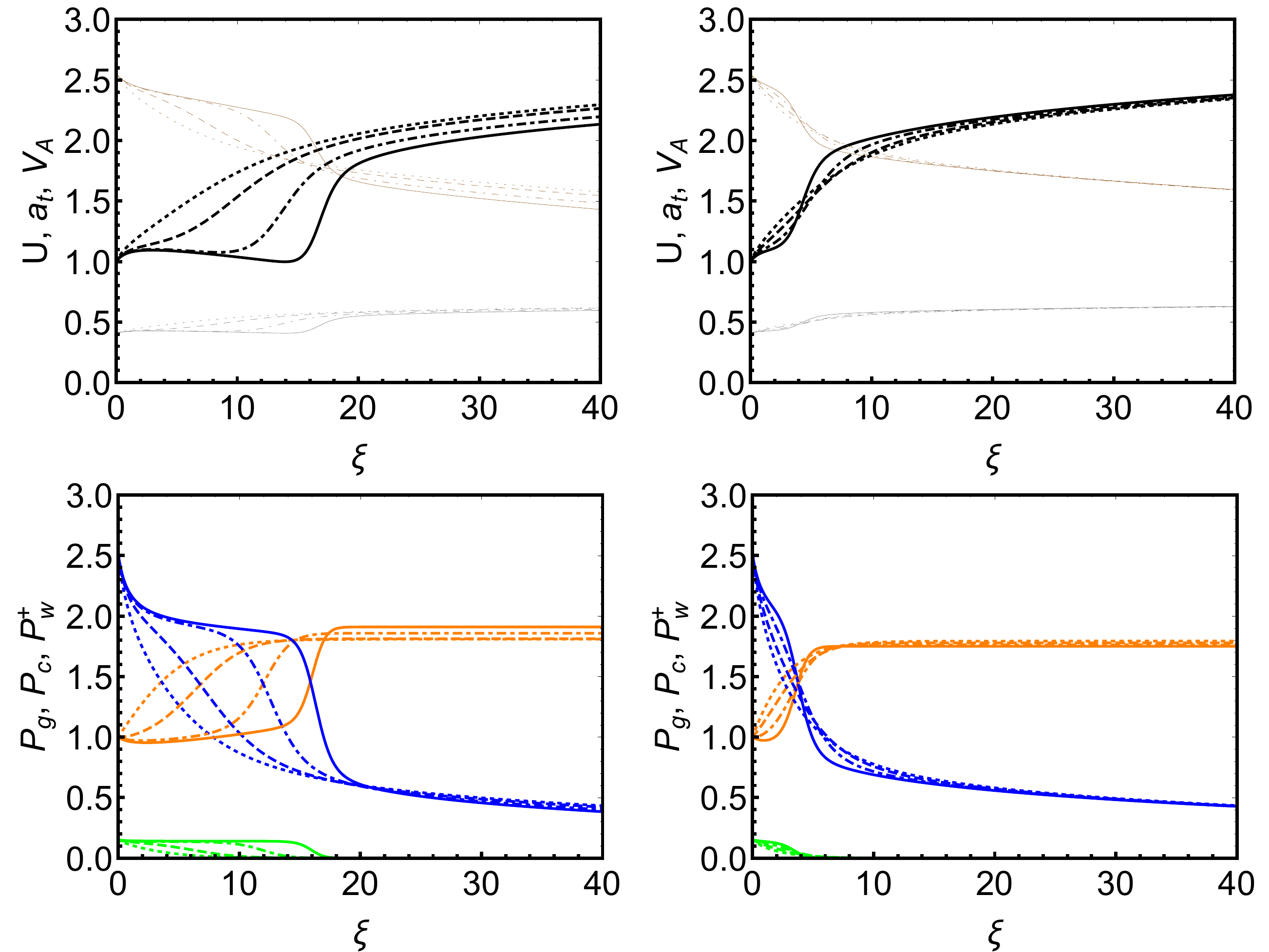}
\vskip 0.5cm
\caption{
Comparison of results with different cosmic ray coupling strength. The cosmic ray diffusion coefficient is defined as $\kappa= \eta/3/P_w^\pm$. Here, we compare solutions for different $\eta$ of the cases with cooling in Figure~\ref{fig:3fluid_diff}. 
{\it Left  column}: Cooling is ``ON’’ and wave damping is ``OFF’’ ($\delta_c=0.1$, $\delta^+=0$).
{\it Right column}: Both cooling and wave damping are ``ON’’ ($\delta_c=0.1$, $\delta^+=0.1$).
Four different cases are shown, (1) $\eta=5$ (dotted line), (2) $\eta=2$ (dashed line), (3) $\eta=1$ (dot-dashed line), and (4) $\eta=0.5$ (solid line).
All other parameters are the same as Figure~\ref{fig:3fluid_diff}, except that $D_{cb}$ is adjusted to produce transonic solutions.
}
\label{fig:3f-diff-eta-1}
\end{figure}

We are also interested in the effect of the cosmic-ray coupling strength on the results with cooling (in which the physically allowable solutions are transonic solutions).
The cosmic ray diffusion coefficient in the three-fluid case is $\kappa= \eta/3/P_w^\pm$. The coupling is stronger for smaller $\eta$. In Figure~\ref{fig:3fluid_diff}, $\eta=1$. For the cases with cooling in Figure~\ref{fig:3fluid_diff}, we work out the corresponding solutions for $\eta=5$, 2, 1, and 0.5 (by adjusting $D_{cb}$), and the result are shown in Figure~\ref{fig:3f-diff-eta-1}. In the figure, the left column is the case with cooling ``ON'' and wave damping ``OFF'' ($\delta_c=0.1$, $\delta^+=0$) and the right column with both cooling and wave damping ``ON'' ($\delta_c=0.1$, $\delta^+=0.1$). The rapid increase in the flow speed and cosmic ray pressure (and decrease in the thermal pressure) become steeper when $\eta$ is smaller. The solutions will be unphysical for even smaller $\eta$. Moreover, we note that for $\eta=0$ (strong coupling limit), the system belongs to Section~\ref{sec:without_diffusion}. The corresponding solution with the same boundary conditions $(U_b,P_{gb},P_{cb},P_{wb}^+,\rho_b,B_b)$ as those in Figure~\ref{fig:3fluid_diff} is unphysical.
%%%%%%%%%%%%%%%%%%%%%%%%%%%%%%%%%%%%%%%%%%%%%%%%%%%%
%%%%%%%%%%%%%%%%%%%%%%%%%%%%%%%%%%%%%%%%%%%%%%%%%%%
%%%%%%%%%%%%
\section{Example of a four-fluid outflow}
\label{sec:4fluid-wdiff}
\begin{figure}[htb]
\centering
\vskip 0.5cm
\includegraphics[width=0.5\textwidth]{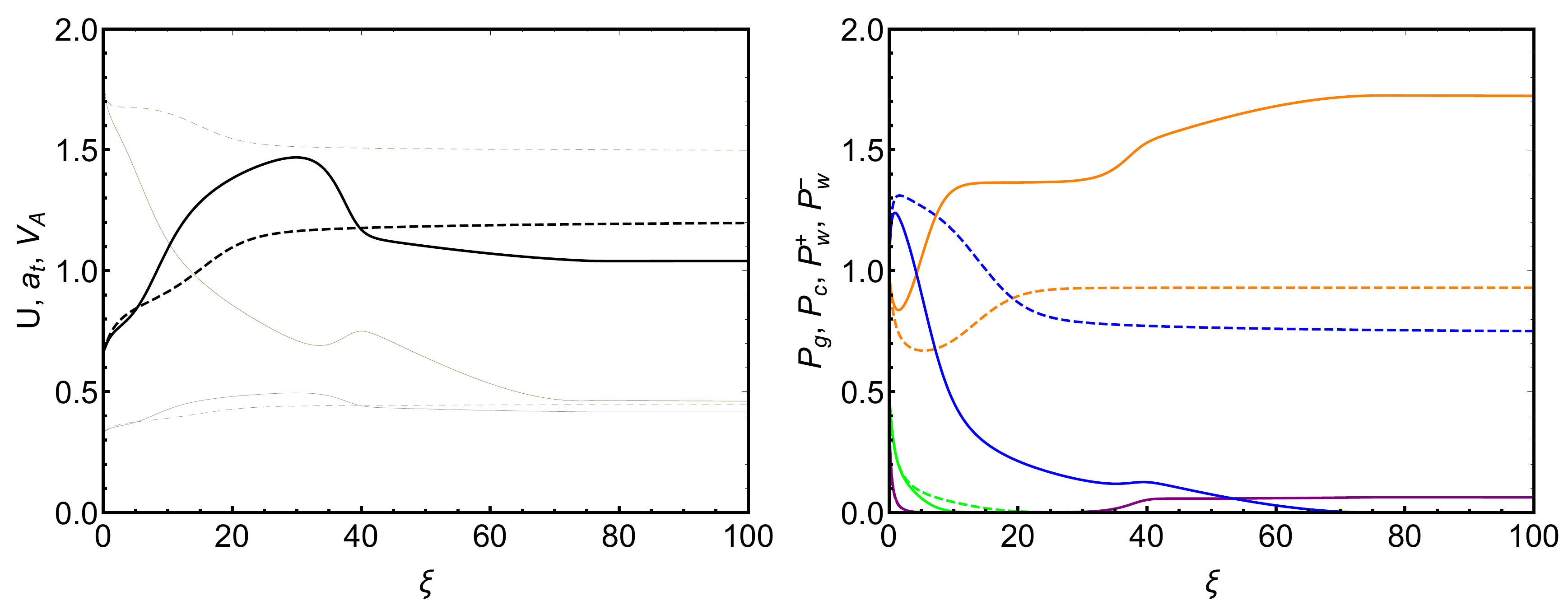}
\vskip 0.5cm
\caption{
Example of transonic and subsonic solutions of a four-fluid system with both forward- and backward-propagating waves. In this figure, $\gamma_g=5/3$, $\gamma_c=4/3$, $GM=1$, $a=1$, $\eta=1$. 
Two cases are compared, (1) 
dashed line: Cooling is “OFF” ($\delta_c=0$) and (2)
solid line: Cooling is “ON” ($\delta_c=0.1$). Wave damping is “ON” for both cases ($\delta^\pm=1$).  
{\it Left panel}: Black, thin brown, and grey lines show the flow speed, the effective sound speed, and the Alfv\'en speed, respectively.
{\it Right panel}: Blue, orange, green, and purple lines show the pressures of thermal gas, cosmic ray, forward-propagating wave, and backward-propagating wave, respectively. The boundary conditions $(U_b,P_{gb},P_{cb},P_{wb}^+,P_{wb}^-,\rho_b,B_b)$ are the same for the two cases. In this example, $U_b=2/3$, $P_{gb}=1$, $P_{cb}=1$, $P_{wb}^+=0.5$, $P_{wb}^-=0.4$, $\rho_b=1$, $B_b=\psi_B^\prime=1/3$. The cosmic ray pressure gradient or diffusive flux at the base is adjusted to get the transonic solution if cooling is “ON” and get a physically allowable subsonic solution if the cooling is “OFF.”
}
\label{fig:4fluid-wdiff}
\end{figure}

The four-fluid model is the most comprehensive hydrodynamic model for cosmic ray-plasma system. It involves both forward- and backward-propagating waves. It is far more complicated than the three-fluid model discussed in Section~\ref{sec:withandwithout_diffusion}. When both waves exist, stochastic acceleration cannot be neglected ($1/\tau> 0$). In addition, it is possible for the cosmic ray streaming instability (represented by the cosmic ray pressure gradient term in Equation~(\ref{eq:two_wave_wave})) to have an opposite effect on the two waves. Here, we present an interesting example to illustrate the intriguing relation between different factors.
In Figure~\ref{fig:4fluid-wdiff}, we compare the case with cooling to the case without cooling. Both cases have the same wave damping and the same boundary conditions $(U_b,P_{gb},P_{cb}, P_{wb}^+, P_{wb}^-,\rho_b,B_b)$ at the base of the gravitational potential well. By adjusting the cosmic ray diffusive flux $D_{cb}$ at the base, we solve the wind equation (Equation~(\ref{eq:two_wave_wind})) together with the corresponding pressure equations (Equations (14)-(16)) for transonic solution for the case with cooling and for subsonic solution that can extend to large distances (i.e., physically allowable solution) for the case without cooling.

The case with cooling has a number of distinct features, as shown in Figure~\ref{fig:4fluid-wdiff}. The velocity profile (black solid line in the left panel of Figure~\ref{fig:4fluid-wdiff}) has an obvious hump. Non-monotonicity is not observed in cases without cooling. The rapid decrease of velocity is related to the (second) increase of cosmic ray pressure (orange solid line in the right panel) and the rise in the pressure of the backward-propagating wave (purple solid line in the right panel). Moreover, the backward-propagating wave survives at large distances because wave damping approaches to zero. Cooling causes the temperature ( $T\propto P_g/\rho$) to decrease. At this point, we recall the working model for cooling ($\Gamma\propto\rho^2\sqrt{P_g/\rho}$) and wave damping ($L_w^\pm\propto P_w^{\pm\,2}\sqrt{P_g/\rho B^2}$). For details, see the end of Section~\ref{sec:model}. When the temperature becomes very low, both cooling and wave damping becomes ineffective.

It is worth pointing out that in this particular example, the wave pressures of the case with cooling are practically zero between $\xi=10$ and $30$. Thus, cosmic ray diffusion coefficient becomes very large and cosmic ray is decoupled from the plasma. However, when the backward-propagating wave rises again (around $\xi=40$), cosmic rays are recoupled to the plasma again.

Another interesting result of this particular example is the flow speed at large distances of the subsonic solution is actually larger than the transonic solution (black dashed and solid lines in the left panel of Figure~\ref{fig:4fluid-wdiff}).
 %%%%%%%%%%%%%%%%%%%%%%%%%%%%%%%%%%%%%%%%%%%%%%%%%%%%%%%%%%%%%
%%%%%%%%%%%%%%%%%%%%%%%%%%%%%%%%%%%%%%%%%%%%%%%%%%%%%%%%%%%%%%
\begin{figure*}[htb]
\centering
\vskip 0.5cm
\includegraphics[width=0.8\textwidth]{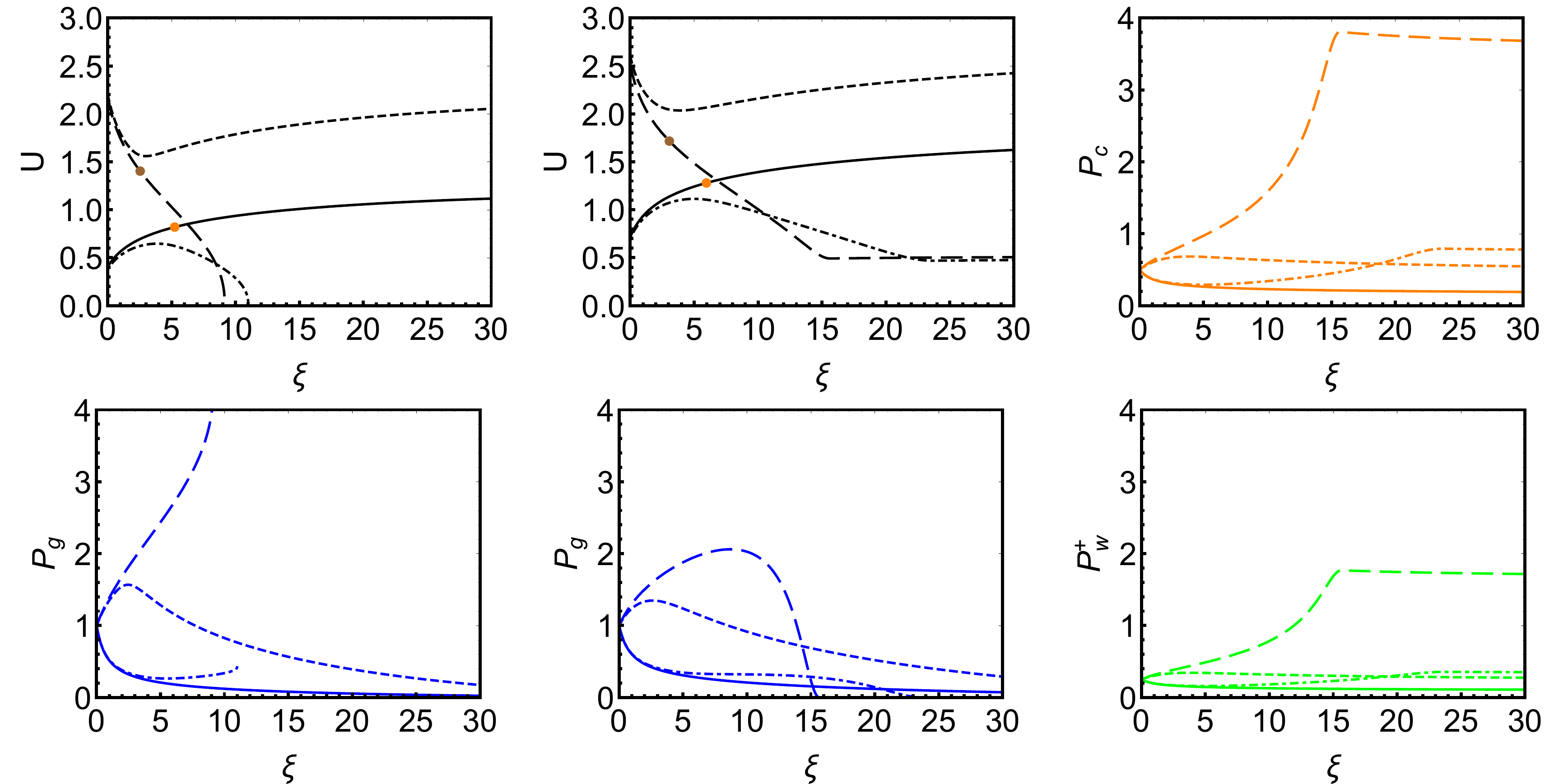}
\vskip 0.5cm
\caption{
Comparison of solutions of a three-fluid system with cooling but without cosmic ray diffusion and wave damping. In this figure, $\gamma_g=5/3$, $\gamma_c=4/3$, $GM=1$, $a=1$, $\delta_c=0.1$, $\delta^+=0$. Four solutions are shown: (1) 
Solid line: Subsonic-supersonic transonic solution; (2) Long-dashed line: Supersonic-subsonic transonic solution; (3) Dot-dashed line: Subsonic solution; and (4) Dashed line: Supersonic solution. The three-fluid case is shown in the middle and right columns, with boundary conditions $(P_{gb},P_{cb},P_{wb}^+,\rho_b,B_b)=(1,0.5,0.25,1,0.25)$. For comparison, a corresponding pure thermal case is shown in the left column with boundary conditions 
$(P_{gb},\rho_b,B_b)$=$(1,1,0.25)$.
In the top-left and top-middle panels, the orange and brown dots are the sonic points of the subsonic-supersonic and supersonic-subsonic transonic solutions, respectively.
}
\label{fig:solutioncurvescompare}
\end{figure*}
%%%%%%%%%%%%%%%%%%%%%%%%%%%%%%%%%%%%%%%%%%%%%%%%%%%%%%%%%%%%%%

\section{Summary and discussion}
\label{sec:Summary}
In this article, we study the role of cosmic rays and waves on outflows with cooling. Our fiducial outflow example is that of galactic winds. For simplicity, we only considered steady-state outflows in a constant flux tube (i.e., one-dimensional flow). For a pure thermal outflow in subsonic regime, cooling causes the outflow speed to decrease towards small values and zero at finite distances.
The density becomes very large at finite distances and it would be difficult to match the density, for instance, in the intergalactic medium. However, we find that if cosmic rays and waves are present and the cosmic ray diffusion can be neglected, then still it is possible to have an outflow to large distances even in the subsonic regime. Let us examine the case of three-fluid system with no cosmic ray diffusion and no wave damping in further detail. In this case, $P_c$ and $P_w^+$ can be expressed in terms of $U$ explicitly (see Equations ~(\ref{eq:pc_no_diff}) and ~(\ref{eq:pwp_no_diff})).
To facilitate discussion, let us write down the equations governing the flow speed, thermal pressure, and temperature as:
\begin{align}\label{eq:wwcooling}
   \left(1- M^{-2}_{\rm eff}\right)U\frac{dU}{d\xi} = -\left(g_e^\dagger - \Gamma^\dagger\right) \,,
\end{align}
\begin{align}\label{eq:thermalgas}
   \left(1- M^{-2}_{\rm eff}\right)\frac{1}{\rho}\frac{dP_g}{d\xi} = M^{-2}_{\rm g} \left(g_e^\dagger - \beta_P \Gamma^\dagger \right)\,,
\end{align}
\begin{align}\label{eq:tempeq}
   \left(1- M^{-2}_{\rm eff}\right)\frac{k_B}{m}\frac{dT}{d\xi} = \frac{(\gamma_g-1)M^{-2}_{\rm g}}{\gamma_g} \left(g_e^\dagger - \beta_T\Gamma^\dagger \right) \,,
\end{align}
where $M_{\rm eff}=U/a_{\rm eff}$, $a_{\rm eff}$ is given by Equation~(\ref{eq:sound_speed_1}), $M_g^{-2}=\gamma_g P_g/\rho U^2$, $g_e^\dagger=-g_e>0$, $\Gamma^\dagger=(\gamma_g-1)\Gamma/\rho U>0$, and 
\begin{align}\label{eq:beta}
  \beta_P = \frac{\left(1- M_{\rm eff}^{-2}+M^{-2}_g\right)}{M^{-2}_g}, \quad \beta_T=\frac{\gamma_g\beta_P-1}{(\gamma_g-1)}\,.
\end{align}
We note that $M_{\rm eff}=M_g$ for pure thermal outflow. In the supersonic regime, $\beta_T>\beta_P>1$, while in the subsonic regime $\beta_T<\beta_P<1$ ($\beta_P$ can be negative, and $\beta_T<0$ if $\beta_P<1/\gamma_g$).

\noindent
In the subsonic regime ($1-M_{\rm eff}^{-2}<0$):
\begin{itemize}
\item (a1): If $\beta_T\Gamma^\dagger$ < $\beta_P\Gamma^\dagger$ < $\Gamma^\dagger$ < $g_e^\dagger$, then $dU/d\xi$ > 0, $dP_g/d\xi$ < 0, $dT/d\xi$ < 0
\item (a2): If $\beta_T\Gamma^\dagger$ < $\beta_P\Gamma^\dagger$ < $g_e^\dagger$ < $\Gamma^\dagger$,  then $dU/d\xi$ < 0, $dP_g/d\xi$ < 0, $dT/d\xi$ < 0
\item (a3): If $\beta_T\Gamma^\dagger$ < $g_e^\dagger$ < $\beta_P\Gamma^\dagger$ < $\Gamma^\dagger$,  then $dU/d\xi$ < 0, $dP_g/d\xi$ > 0, $dT/d\xi$ < 0
\item (a4): If $g_e^\dagger$ < $\beta_T\Gamma^\dagger$ < $\beta_P\Gamma^\dagger$ < $\Gamma^\dagger$, then $dU/d\xi$ < 0, $dP_g/d\xi$ > 0, $dT/d\xi$ > 0
\end{itemize}
In the supersonic regime ($1-M_{\rm eff}^{-2}>0$):
\begin{itemize}
\item (b1): If $\Gamma^\dagger$ < $\beta_P\Gamma^\dagger$ < $\beta_T\Gamma^\dagger$ < $g_e^\dagger$, then $dU/d\xi$ < 0, $dP_g/d\xi$ > 0, $dT/d\xi$ > 0
\item (b2): If $\Gamma^\dagger$ < $\beta_P\Gamma^\dagger$ < $g_e^\dagger$ < $\beta_T\Gamma^\dagger$, then $dU/d\xi$ < 0, $dP_g/d\xi$ > 0, $dT/d\xi$ < 0
\item (b3): If $\Gamma^\dagger$ < $g_e^\dagger$ < $\beta_P\Gamma^\dagger$ < $\beta_T\Gamma^\dagger$, then $dU/d\xi$ < 0, $dP_g/d\xi$ < 0, $dT/d\xi$ < 0
\item (b4): If $g_e^\dagger$ < $\Gamma^\dagger$  < $\beta_P\Gamma^\dagger$ < $\beta_T\Gamma^\dagger$, then $dU/d\xi$ > 0, $dP_g/d\xi$ < 0, $dT/d\xi$ < 0
\end{itemize}
We note that a system can exist in a range (a4) of the subsonic regime only for a small extent of $\xi$ (if at all). In range (a4), as $\xi$ increases, $U$ decreases, and $P_g$ increases, which implies that $\beta_P$ decreases. When $\beta_P$ is less than $1/\gamma_g$, the value of $\beta_T$ is negative and the system will leave the current range (a4) and enter range (a3). 

In the subsonic regime when cooling is present, the flow speed decreases as $\xi$ increases if the system is in range (a2), (a3) or (a4).
To prevent the flow from stalling at finite distance, the cooling has to be turned off before this happens. 
The system will then stay in range (a1), and the flow speed will gradually increase towards a finite value as $g_e\approx 0$ at large distances. Suppose the cooling function vanishes as temperature goes to a small value (e.g., the cooling function $\Gamma\propto \rho^2 \sqrt{T}$ approaches zero as $T$ approaches zero).
When the flow is in the range of (a2) or (a3) of the subsonic regime, it has a tendency to become stalled. In these ranges, the temperature also decreases. When the temperature decreases rapidly enough, the cooling can be turned off before the speed of the flow reaches zero.
It is then possible to have a physically allowable solution that extends to large distances. 
We then consider what the role of cosmic rays and waves is on the flow in the subsonic regime. We thus denote $\beta^{\rm therm}_p$ and $\beta^{\rm therm}_T$ for the pure thermal case and $\beta^{\rm 3f}_p$ and $\beta^{\rm 3f}_T$ for the three-fluid case (see Equation~(\ref{eq:beta}) for the definition of $\beta_P$ and $\beta_T$). If $U$ and $P_g$ are the same in the two cases, then $\beta_P^{\rm therm}>\beta_P^{\rm 3f}$ and $\beta_T^{\rm therm}>\beta_T^{\rm 3f}$. 
When $\beta_T$ is smaller, the temperature decreases more rapidly. Hence, the presence of cosmic rays and waves helps the temperature to drop faster and turns off the cooling. As a result, it is easier to have a subsonic flow when cosmic rays and waves are included.

Figure~\ref{fig:solutioncurvescompare} illustrates the effect of cosmic rays and waves on outflows, in particular, subsonic flows. 
Radiative cooling by bremsstrahlung ($\Gamma\propto\rho^2\sqrt{T}$) is adopted here. The left column shows the pure thermal case. The middle and right columns show the three-fluid case (without cosmic ray diffusion and wave damping). The figure shows four types of solutions, subsonic, subsonic-supersonic, supersonic-subsonic, and supersonic. 
All the given solutions have the same boundary conditions $(P_{gb} ,\rho_b,B_b)$ for both cases and, in addition, the same $(P_{cb},P_{wb}^+)$ in the case of three-fluid systems. 
Also, $U_b$ is adjusted to get the two transonic solutions. $U_b$ for the subsonic (supersonic) solution is slightly smaller (larger) than the one for the subsonic-supersonic (supersonic-subsonic) transonic solution. 
As shown in the top-left panel of the figure for the pure thermal outflow, the subsonic (dot-dashed line) and the supersonic-subsonic (long dashed line) solutions are not capable of being extended to large distances and the flow velocity goes to zero at finite distance.
The top-middle panel shows that with the help of cosmic rays and waves, the temperature drops to small value and the cooling is off before the flow velocity goes to zero. The flow is then able to enter range (a1) and approaches a finite flow speed at large distances (see middle column of Figure~\ref{fig:solutioncurvescompare}.
The sudden change in $U$ (and $P_c$, $P_w^+$) in the subsonic region occurs when the temperature drops to zero and the cooling function also becomes zero. In this particular case, the flow goes from range (a2) to range (a1). 

The orange dot in the top-left and top-middle panels of Figure 4 is the sonic point of the subsonic-supersonic transonic solution, i.e., the location where the flow changes from subsonic to supersonic. At this point $\beta_T=\beta_P=1$, and the flow moves from range (a1) in the subsonic regime to range (b4) in the supersonic regime. The brown dot is the sonic point of the supersonic-subsonic transonic solution, and the flow moves from range (b1) in the supersonic regime to range (a4) in the subsonic regime (and leaves range (a4) soon). The apparent crossing of the two transonic solutions in the figure is only a projection effect.

We conclude that in the case of three-fluid system without cosmic ray diffusion and wave damping, a subsonic outflow is possible provided that the amount of cosmic ray and wave is large enough. We caution that if cosmic ray diffusion is allowed, the line of reasoning above does not apply and the existence of subsonic flow will be in question (we have tried a range of parameters for physically allowable subsonic solutions but without success). 

As discussed above and shown in Figure~\ref{fig:solutioncurvescompare}, a three-fluid subsonic outflow is possible. However, the flow speed at large distances is small, that is, the density is large (e.g., the example in Figure~\ref{fig:solutioncurvescompare} has a density at large distances larger than at the base of the potential well). We are thus more interested in subsonic-supersonic transonic solutions.

In summary, in this article, we present a study of the steady-state subsonic-supersonic transonic outflow of cosmic-ray plasma systems with cooling. We only considered one-dimensional flow. Our fiducial model is based on the galactic wind. We choose bremsstrahlung radiation as working model for cooling and nonlinear Landau damping for wave damping.

We examined the three-fluid model or one-wave model (which comprises thermal plasma, cosmic ray, and self-excited forward-propagating Alfv\'en wave) without cosmic ray diffusion in detail (Section~\ref{sec:without_diffusion}).
Here $(P_{g},P_{c}, P_{w}^+,\rho,B)$ are set at the base of the gravitational potential well. Here, the value of $U$ at the base is adjusted to obtain a transonic solution. With the same pressures, density, and magnetic field at the base, the profiles of the flow speed and pressures of the cases with and without wave damping are rather similar. However, even with the same thermal pressure and density at the base, the flow speed of the pure thermal transonic outflow is significantly lower than the three-fluid flow, which is expected, as the effective sound speeds are different for both cases (see Figure~\ref{fig:3-fluid-nodiff}).

When cosmic ray diffusion is taken into consideration in the three-fluid system, the behavior exhibited by the solutions can be quite different, particularly when the cosmic-ray diffusion coefficient depends on waves (Section~\ref{sec:3fluid-wdiff}). In this case, $(U,P_{g},P_{c}, P_{w}^+,\rho,B)$ are set at the base and the cosmic ray pressure gradient or diffusive flux ($D_{cb}$) is adjusted to get transonic solutions. We compare
the case of without both cooling and wave damping to cases of either cooling or wave damping, or both.
In this example, cases with either cooling or wave damping show ``rapid change’’ in velocity and pressure profiles.
This is a characteristic feature of quasi-thermal outflows (Ramzan et al. 2020; see Appendix~\ref{sec:Behaviorofpressures}).
When cooling is considered, subsonic-supersonic transition is the only physical solution.
In the presence of wave damping, the ``sonic point’’ seems to be pulled closer to the base (compare the middle and the right columns of Figure~\ref{fig:3fluid_diff}).
When the coupling between cosmic rays and the thermal gas increases ($\eta$ decreases), the ``rapid change’’ in the profiles becomes ``steeper’’ (see Figure~\ref{fig:3f-diff-eta-1}). 
In this example, there will be no physical solution for a small-enough value for $\eta$.

In Section~\ref{sec:4fluid-wdiff}, we gave an example on the most comprehensive cosmic ray-MHD system, the four-fluid model or two-wave model (where both forward- and backward-propagating waves are included). Due to complicated interactions between different components, the profiles of the flow and pressures may show some interesting features.
For instance, in this particular example, when cooling is ``ON’’ (wave damping is also ``ON’’) the flow speed has a hump that is not observed in cases without cooling, Moreover, along the flow cosmic ray is first coupled to the plasma, subsequently decoupled and then recoupled again at the further down the flow (indicated by the amount of waves along the flow, shown as green and purple solid lines in the right panel of Figure~\ref{fig:4fluid-wdiff}).
In addition, in this particular example, the flow speed at large distances of the subsonic solution can be larger than that of the subsonic-supersonic transonic solution (left panel of Figure~\ref{fig:4fluid-wdiff}).

\begin{acknowledgements}
CMK and BR are supported in part by the Taiwan Ministry of Science and Technology grant MOST 109-2112-M-008-005 and MOST 110-2112-M-008-005.
\end{acknowledgements}

\bibliographystyle{aa}

\appendix
\section{Critical behaviors of three-fluid systems}
\label{sec:critical_behaviour}
In this appendix, we analyse the critical behaviour (or fixed points) of two three-fluid systems. This is important for the study of transonic solutions (i.e., solutions passing through fixed points or critical points).
\subsection{Three-fluid system with cooling but without cosmic ray diffusion and wave damping}
\label{sec:3f_without_diffusion}
\begin{figure*}[htb]
\centering
\vskip 0.1cm
\includegraphics[width=0.8\textwidth]{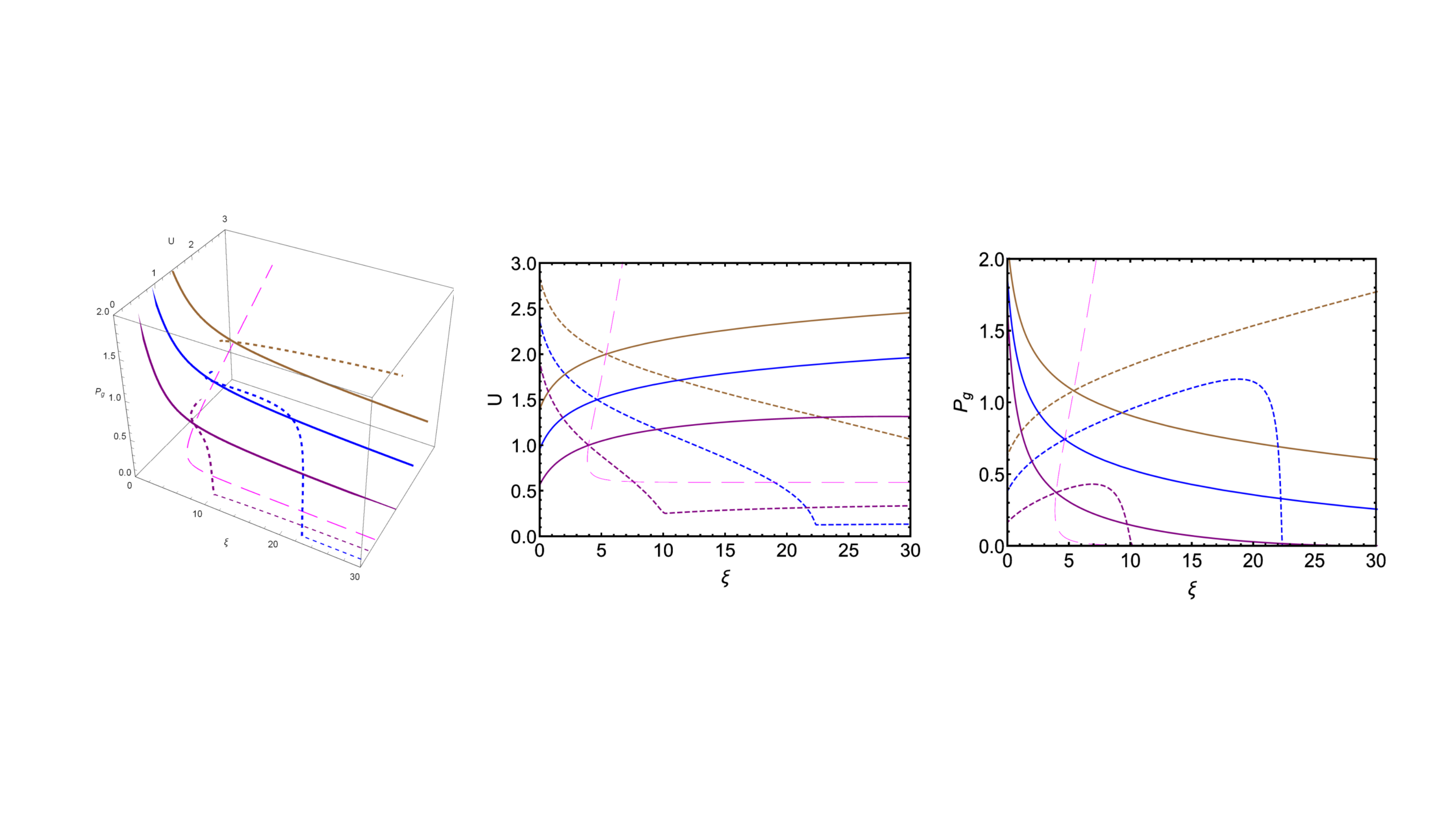}
\vskip 0.1cm
\caption{
Critical line and three sets of transonic solutions of a three-fluid system without cosmic ray diffusion.
In this figure, $\gamma_g=5/3$, $\gamma_c=4/3$, $GM=1$, $a=1$,
$\psi^\prime_B=1/2$, $\psi^\prime_m=1$, $\delta_c=0.1$, $A_c=0.5$, $A_w^+=0.25$.
Magenta long-dashed line is the critical line.
Solid lines: subsonic-supersonic transitions. Dashed lines: supersonic-subsonic transitions.
}
\label{fig:3-fluid-CL}
\end{figure*}

We considered a three-fluid system (one-wave system with forward-propagating wave) without cosmic ray diffusion and wave damping. The cosmic ray pressure and wave pressure can be expressed in terms of flow speed by solving Equations~(\ref{eq:two_wave_wave}) and ~(\ref{eq:two_wave_cr}) (see Ramzan et al. 2020):
\begin{equation}\label{eq:pc_no_diff}
  P_c = A_c \left|\frac{\rho}{(1+ \tilde{\psi}  \sqrt{\rho})}\right|^{\gamma_c} \,,
\end{equation}
\begin{equation}\label{eq:pwp_no_diff}
  P^+_w = \left[A^+_w -\frac{\gamma_c P_c}{2(\gamma_c-1)}\frac{\tilde{\psi}  \sqrt{\rho})}{\left(1+ \tilde{\psi}  \sqrt{\rho})\right)}\right]
  \frac{\rho^{3/2}}{\left(1+ \tilde{\psi}  \sqrt{\rho})\right)^2}\,,
\end{equation}
where
$\tilde{\psi}= \psi^\prime_B/(\sqrt{\mu_0} \psi^\prime_m)$, \quad $\rho=\psi^\prime_m/U$.
The governing set of equations can be written as a three-dimensional autonomous system,
\begin{equation}\label{eq:matric_CL}
 \frac{d}{ds} \begin{bmatrix} \xi \\U\\\ P_g \end{bmatrix}=
\begin{bmatrix} \mathcal{P}\\\mathcal{Q}\\\mathcal{R} \end{bmatrix},
\end{equation}
where
\begin{align}\label{eq:Critical_Line}
\mathcal{P} = \left(1- M_{\rm eff}^{-2}\right),
\end{align}
\begin{align}\label{eq:Critical_Line_1}
 \mathcal{Q} = \frac{1}{U} \left[g_{e} + \frac{(\gamma_g-1)\Gamma}{\psi^\prime_m}  \right],
\end{align}
\begin{align}
 \mathcal{R} = -\frac{\gamma_g p_g}{U} \mathcal{Q}   - \frac{(\gamma_g-1)\Gamma}{U} \mathcal{P}  \,,
\end{align}
where  $M_{\rm eff}$= $U/a_{\rm eff}$, $a_{\rm eff}^{2}$ is given by Equation~(\ref{eq:sound_speed_1}).
The fixed point or critical point of the system is determined by setting the right hand side of the Equation~(\ref{eq:matric_CL}) equal to zero.
 This can be achieved by:
 \begin{align}
\mathcal{P} = 0,  \quad\quad \mathcal{Q} = 0,
\end{align}
which gives a critical line in the three-dimensional ($\xi$,$U$,$P_g$) space,
\begin{align}\label{eq:B11}
 U_s = \mu, 
 \end{align}
\begin{equation}\label{eq:C_L}
 P_{gs}  = \frac{1}{\gamma_g} \left [\psi^\prime_m U_s-\gamma_c P_{cs}\frac{(1+ \frac{1}{2}\tilde{\psi}  \sqrt{\rho_s})^2 }{(1 + \tilde{\psi}  \sqrt{\rho_s})^2}-\frac{3}{2}P^+_{ws}\frac{(1+ \frac{1}{3}\tilde{\psi}  \sqrt{\rho_s}) }{(1 +\tilde{\psi}  \sqrt{\rho_s})} \right] \,,
\end{equation}
\begin{equation}\label{eq:sonicline}
\xi_s =\sqrt{\frac{GM \psi^\prime_m}{(\gamma_g-1) \Gamma_s}} - a_K,
\end{equation}
where $\mu$ is the parametric variable of the critical line and $\Gamma_s=\Gamma_s(\rho_s,P_{gs})$. 
For illustration, we take $\Gamma=\delta_c \rho^2(P_g/\rho)^{1/2}$ and show the result in Figure~\ref{fig:3-fluid-CL}. The critical line is the long-dashed pink line. Linear analysis show that the points on the critical line are saddle-points except for those close to $P_g=0$. In the figure, we show transonic solutions from three points on the critical line. For each point, there are two transonic solutions: one from subsonic to supersonic (solid line) and the other from supersonic to subsonic (dashed line).
%%%%%%%%%%%%%%%%%%%%%%%%%%%%%%%%%%%%%%%%%%%%%%%%%%%%%%%%%%%%%
%%%%%%%%%%%%%%%%%%%%%%%%%%%%%%%%%%%%%%%%%%%%%%%%%%%%%%%%%%%%%%
\subsection{Three-fluid system with cosmic ray diffusion but without cooling and wave damping}
\label{sec:3f_with_diffusion_withoutcooling}
We also considered a three-fluid system (with forward-propagating wave) in which cosmic ray diffusion is included but cooling and wave damping are neglected. The diffusion coefficient is inversely proportional to the wave pressure.
In addition to magnetic flux and mass flux, the total energy flux and the wave-action integral are constant in this system:
\begin{align}\label{eq:total_energy_intergral}
F^\prime_{\rm tot} = \frac{1}{2} U^2+F^\prime_{\rm g}+F^\prime_{\rm c}+F^{\prime +}_w +\Psi_e \,,
\end{align}
\begin{align}\label{eq:wave_action_intergral}
\mathcal{W}_A^\prime= \left[F^\prime_{\rm c}+\frac{2P^+_{w}U}{\psi^\prime_m}\frac{(1+ \tilde{\psi}  \sqrt{\rho})^2}{\tilde{\psi}\sqrt{\rho}}\right] \,,
\end{align}
$F^\prime_g=F_g/\psi^\prime_m$, $F^\prime_c=F_c/\psi^\prime_m$, $F^{\prime+}_w=F_w^+/\psi^\prime_m$ and $F_g$, $F_c$ and $F_w^+$ are given by Equations~(\ref{eq:gflux}), ~(\ref{eq:crflux}), and (9). Here $\Psi_e$ is the external gravitational potential and:
\begin{equation}\label{eq:def_Pg}
\rho U=\psi^\prime_m,\quad P_g=A_g \rho^{\gamma_g},\quad \kappa=\eta/3P^+_w.
\end{equation}
Solving Equations~(\ref{eq:total_energy_intergral}) and ~(\ref{eq:wave_action_intergral}) for $P^+_w$ and $F_c^\prime$,
\begin{align}\label{eq:total_WP}
 P^{+}_{w}= & \,\frac{\tilde{\psi}\rho^{3/2}}{2(1+ \frac{1}{2} \tilde{\psi}\sqrt{\rho})} \left[\frac{1}{2} U^2+\frac{\gamma_g P_g}{(\gamma_g-1)\rho}+\Psi_e-F^\prime_{\rm tot}+\mathcal{W}_A^\prime\right] \,,
\end{align}
\begin{align}\label{eq:total_CRF}
 F_{c}^\prime=& \frac{(1+ \tilde{\psi}\sqrt{\rho})^2}{(1+ \frac{1}{2} \tilde{\psi}\sqrt{\rho})}
  \left[ F^{\prime}_{\rm tot}-\frac{1}{2} U^2-\frac{\gamma_g P_g}{(\gamma_g-1)\rho}-\Psi_e\right] \nonumber \\   
  &\quad\quad -\frac{3\tilde{\psi}\sqrt{\rho}(1+\frac{2}{3} \tilde{\psi}\sqrt{\rho})}{2(1+\frac{1}{2} \tilde{\psi}\sqrt{\rho})}\mathcal{W}_A^\prime\ \,,
\end{align}
The governing set of equations can be written as a set of three-dimensional autonomous systems:
\begin{equation}\label{eq:Spoint}
 \frac{d}{ds} \begin{bmatrix} \xi \\U\\\ P_c \end{bmatrix}=
\begin{bmatrix} \mathcal{P}\\\mathcal{Q}\\\mathcal{R} \end{bmatrix},
\end{equation}
where
\begin{align}\label{eq:B1}
\mathcal{P} = \left(1- M_{\rm eff}^{-2}\right), 
 \end{align}
\begin{align}\label{eq:B2}
  \mathcal{Q} = & \, \frac{g_e}{U}-  \mathcal{D}  \,,
\end{align}
\begin{align}\label{eq:B_2}
  \mathcal{R} = & \,\frac{(1+ \tilde{\psi}\sqrt{\rho})}{(1+ \frac{1}{2}\tilde{\psi}\sqrt{\rho})} \psi^\prime_m\mathcal{DP}  \,,
\end{align}
\begin{align}\label{eq:B3}
 \mathcal{D} = & \frac{(\gamma_c-1)}{\kappa} (1+ \tilde{\psi}\sqrt{\rho})\nonumber \\
  & * \left[ \frac{1}{2} \right.  U^2+\frac{\gamma_g P_g}{(\gamma_g-1)\rho}+\frac{\gamma_c P_c}{(\gamma_c-1)\rho}\frac{(1+ \frac{1}{2}\tilde{\psi}\sqrt{\rho})}{(1+ \tilde{\psi}\sqrt{\rho})}\nonumber \\
  & +\Psi_e-F^{\prime}_{\rm tot}  \left. +\frac{3\tilde{\psi}\sqrt{\rho}(1+\frac{2}{3} \tilde{\psi}\sqrt{\rho})}{2(1+ \tilde{\psi}\sqrt{\rho})^2}\mathcal{W}_A^\prime \right]\,,
\end{align}
where  $M_{\rm eff}$= $U/a_{\rm eff}$, $a_{\rm eff}^{2}$ is given by Equation~(\ref{eq:sound_speed}) with $P_w^-=0$.
 The fixed point or critical point of the system is determined by setting the right hand side of Equation~(\ref{eq:Spoint}) equal to zero.
 This can be achieved via:
  \begin{align}
\mathcal{P} = 0,\quad \quad \mathcal{Q} = 0,
\end{align}
which gives a critical line in three dimensional  ($\xi$,U,$P_c$) space,
 \begin{align}
 \label{eq:B7}
 \Psi_{es} = &\left[F^\prime_{\rm{tot}}-\mathcal{W}_A^\prime -\frac{1}{2} U^2_s-\frac{\gamma_g P_{gs}}{(\gamma_g-1)\rho_s}\right] \nonumber \\
 & +\frac{4(1+\frac{1}{2} \tilde{\psi}\sqrt{\rho_s})(1+ \tilde{\psi}\sqrt{\rho_s})}{3 \tilde{\psi}\sqrt{\rho_s}(1+\frac{1}{3} \tilde{\psi}\sqrt{\rho_s})}\left(U^2_s-\frac{\gamma_g P_{gs}}{\rho_s}\right) ,
 \end{align}
  \begin{align}\label{eq:B8}
 U_s = \mu, 
 \end{align}
   \begin{align}\label{eq:B9}
 \frac{\gamma_c P_{cs}}{(\gamma_c-1)\rho_s} =&
  \frac{\kappa_s  g_{es}}{(\gamma_c-1)U_s(1+\frac{1}{2} \tilde{\psi}\sqrt{\rho_s})} \nonumber \\
 & - \frac{4(1+ \tilde{\psi}\sqrt{\rho_s})^2}{3\tilde{\psi}\sqrt{\rho_s}(1+ \frac{1}{3}\tilde{\psi}\sqrt{\rho_s})}\left(U^2_s-\frac{\gamma_g P_{gs}}{\rho_s}\right) \nonumber \\
  &+\frac{\mathcal{W}_A^\prime}{(1+ \tilde{\psi}\sqrt{\rho_s})}.
 \end{align}
 where $\Psi_{es}=\Psi_e(\xi_{es})$, $g_{es}=g_e(\xi_{es})$, $\kappa_s=\kappa(P^+_{ws})$ and $\rho_s$, $P_{gs}$ and $P^+_{ws}$ are given by Equations~(\ref{eq:def_Pg}) and ~(\ref{eq:total_WP}); also, $\mu$ is the parametric variable of the critical line.
 \section{Summary of three-fluid outflows}
\label{sec:Behaviorofpressures}
In this appendix, we discuss some general properties of the three-fluid system with a forward-propagating wave and cosmic-ray diffusion.
In the realm of physically allowable solutions starting in the subsonic region at the base of potential, the profiles of $U$, $P_c$, and $P_g$ can be categorized into two regimes \citep{Ramzan_2020}: regime A (called cosmic ray accompanied outflows) and regime B (quasi-thermal outflows).
In regime A, the value of $U$ increases ($P_c$ and $P_g$ decreasing) gradually to a constant value at large distances (e.g., Figure~\ref{fig:3-fluid-nodiff}). Cosmic rays are coupled to the thermal gas at all distances (i.e., $P_w^+$ remains finite all the way).
In regime B, values of $U$ and $P_c$ have a general trend of increasing towards a finite value at large distances ($P_g$ decreases).
The profile of $U$ and $P_c$ has a characteristic rapid increase or a ``jump’’ at some distance from the base ($P_g$ has a rapid decrease or a ``drop,’’ e.g., Figure~\ref{fig:3fluid_diff}).
Also, $P_w^+$ decreases monotonically and has a ``drop’’ at the same location and approaches zero afterward. Cosmic rays are decoupled from the thermal gas after the ``jump’’ and $P_c$ becomes constant afterward. Wave damping and cooling are not necessary conditions for the existence of the two regimes, although they may affect parameters defining the regimes. 
In regime A, the set of boundary conditions at the base that gives a physically allowable solution is ``isolated’’ in the sense that any slight change in one (and only one) parameter will push the system into the unphysical domain (however, if more than one parameter can be changed, then it is possible to get into some other physically allowable domain). In contrast, in regime B, a small change in one (and only one) parameter of a physically allowable set may also give physical solutions. This can be understood as follows.

To simplify the argument, let us consider a basic case (system without wave damping and cooling). In this case, only the first two terms on the right hand side of Equation (11) remain (with $e_+=1$ and $e_-=0$), and the second and third terms on the right hand side of Equation (16) vanish. Equation (16) can be viewed as an equation for the cosmic ray diffusive flux $D_c=\kappa dE_c/d\xi$ ($\kappa=\eta/3/P_w^+$). When $M_{\rm eff}<1$ (``subsonic’’ region), Equation (11) and Equation (16) form a positive feedback loop in the following sense. If both $dU/d\xi$ and $D_c$ are negative (positive), then both will be more negative (positive) downstream. There are two possible routes, route (a) and route (b), for avoiding divergence and ending up with a physically allowable solution. 

Let the solution start at the base with $D_c<0$ and $dU/d\xi>0$. $D_c$ then increases while $dU/d\xi$ decreases.
For one particular set of boundary values, $D_c$ becomes less and less negative and approaches to zero at large distances (and $dU/d\xi$ approaches zero from the positive side). This is route (a).

If a solution starts with the same boundary values as route (a) except that $D_c$ is (slightly) smaller, then $dU/d\xi$ will turn negative before $D_c$ goes to zero.
The positive feedback loop activates and the solution will go to the unphysical domain ($U$ and $P_c$ become negative).
However, if a solution starts with the same boundary values as route (a) except that $D_c$ is larger, the situation will be different.

Supposing the solution starts from $D_c<0$ and $dU/d\xi>0$, $D_c$ increases and becomes positive at some finite $\xi$.
The positive feedback loop kicks in and both $D_c$ and $dU/d\xi$ increase rapidly. 
A positive $D_c$, namely, a positive cosmic-ray pressure gradient, will cause $P_w^+$ to diminish via the cosmic-ray streaming instability. 
Cosmic rays will be decoupled from the thermal gas and the outflow behaves like a thermal outflow. If the velocity and thermal pressure at that location belong to the physically allowable region of a pure thermal outflow, then a physically allowable solution is obtained: this is route (b). The set of parameters that allows for route (b) belonging to regime B (quasi-thermal). If the velocity and thermal pressure after the ``rapid change’’ do not belonging to any physically allowable region of a pure thermal outflow, then the set of parameters only allows for route (a) and it belongs to regime A mentioned above.

Once again, the characteristics of a quasi-thermal outflow is a ``jump’’ in velocity and cosmic-ray pressure (and a ``drop’’ in thermal pressure) at the location where $P_w^+$ diminishes towards zero.
A small change in $D_c$ at the base will change the location of the ``jump.’’ Smaller $D_c$ (more negative) pushes the ``jump’’ to larger distances. We point out that there is a minimum $D_c$ such that the ``jump’’ occurs at ``infinity’’ (e.g., the left column of Figure~\ref{fig:3fluid_diff}). This particular solution coincides with route (a). The solutions become unphysical for value of $D_c$ that are smaller than the minimum. 
%%%%%%%%%%%%%%%%%%%%%%%%%%%%%%%%%%%%%%%%%%%%%%%%%%%%%%%%%%%%%%
\end{document}